\documentclass[twocolumn]{aastex631}
\usepackage{graphicx}
\graphicspath{{figures/}}
\usepackage[]{hyperref}
\usepackage{xcolor}
\usepackage{float}
\usepackage{comment}
\let\tablenum\relax \usepackage{siunitx}
\DeclareSIUnit \jansky {Jy}
\usepackage{transparent}
\usepackage{makecell}
\usepackage{multirow}
\newcommand{\Swift}{\textit{Swift}}

\begin{document}

\title{A millimeter rebrightening in GRB 210702A}

\correspondingauthor{Simon de Wet}
\email{DWTSIM002@myuct.ac.za}

\author[0000-0003-2449-1329]{Simon de Wet}
\affiliation{University of Cape Town Astronomy Department, Private Bag X3, Rondebosch, 7701, South Africa}

\author[0000-0003-1792-2338]{Tanmoy Laskar}
\affiliation{Department of Physics \& Astronomy, University of Utah, Salt Lake City, UT 84112, USA}
\affiliation{Department of Astrophysics/IMAPP, Radboud University, PO Box 9010, 6500 GL Nijmegen, The Netherlands}

\author[0000-0002-4488-726X]{Paul J. Groot} 
\affiliation{University of Cape Town Astronomy Department, Private Bag X3, Rondebosch, 7701, South Africa}
\affiliation{Department of Astrophysics/IMAPP, Radboud University, PO Box 9010, 6500 GL Nijmegen, The Netherlands}
\affiliation{South African Astronomical Observatory, PO Box 9, 7935 Observatory, South Africa}

\author[0000-0002-5565-4824]{Rodolfo Barniol Duran}
\affiliation{Department of Physics and Astronomy, California State University, Sacramento, 6000 J Street, Sacramento, CA 95819-6041, USA}

\author[0000-0002-9392-9681]{Edo Berger}
\affiliation{Center for Astrophysics ${\rm \mid}$ Harvard {\rm \&} Smithsonian, 60 Garden St, Cambridge, MA 02138, USA}

\author[0000-0003-3460-506X]{Shivani Bhandari}
\affiliation{ASTRON, Netherlands Institute for Radio Astronomy, Oude Hoogeveensedijk 4, 7991~PD Dwingeloo, The Netherlands}
\affiliation{Joint institute for VLBI ERIC, Oude Hoogeveensedijk 4, 7991~PD Dwingeloo, The Netherlands}
\affiliation{Anton Pannekoek Institute for Astronomy, University of Amsterdam, Science Park 904, 1098~XH Amsterdam, The Netherlands}
\affiliation{CSIRO Space and Astronomy, Australia Telescope National Facility, PO Box 76, Epping, NSW~1710, Australia}

\author[0000-0003-0307-9984]{Tarraneh Eftekhari}
\altaffiliation{NHFP Fellow}
\affiliation{Center for Interdisciplinary Exploration and Research in Astrophysics (CIERA) and Department of Physics and Astronomy, Northwestern University, Evanston, IL 60208, USA}

\author[0000-0001-6869-0835]{C.~Guidorzi}
\affiliation{Department of Physics and Earth Science, University of Ferrara, Via Saragat 1, I-44122 Ferrara, Italy}
\affiliation{INFN -- Sezione di Ferrara, Via Saragat 1, 44122 Ferrara, Italy}
\affiliation{INAF -- Osservatorio di Astrofisica e Scienza dello Spazio di Bologna, Via Piero Gobetti 101, 40129 Bologna, Italy}

\author[0000-0001-7946-4200]{Shiho Kobayashi}
\affiliation{Astrophysics Research Institute, Liverpool John Moores University, 146 Brownlow Hill, Liverpool L3 5RF, UK}

\author[0000-0001-8472-1996]{Daniel A. Perley}
\affiliation{Astrophysics Research Institute, Liverpool John Moores University, 146 Brownlow Hill, Liverpool L3 5RF, UK}

\author[0000-0002-1084-3656]{Re'em Sari}
\affiliation{Racah Institute of Physics, The Hebrew University of Jerusalem, 91904, Israel}
\affiliation{Theoretical Astrophysics, California Institute of Technology, 350-17, Pasadena, CA 91125}

\author[0000-0001-9915-8147]{Genevieve~Schroeder}
\affiliation{Center for Interdisciplinary Exploration and Research in Astrophysics and Department of Physics and Astronomy, Northwestern University, 2145 Sheridan Road, Evanston, IL 60208-3112, USA}

\begin{abstract}
We present X-ray to radio frequency observations of the bright long gamma-ray burst GRB 210702A. Our ALMA 97.5~GHz observations show a significant rebrightening by a factor of $\approx2$ beginning at 8.2 days post-burst and rising to peak brightness at 18.1 days before declining again. This is the first such rebrightening seen in a millimeter afterglow light curve. A standard forward shock model in a stellar wind circumburst medium can explain most of our X-ray, optical and millimeter observations prior to the rebrightening, but significantly over-predicts the self-absorbed radio emission, and cannot explain the millimeter rebrightening. We investigate possible explanations for the millimeter rebrightening and find that energy injection or a reverse shock from a late-time shell collision are plausible causes. Similar to other bursts, our radio data may require alternative scenarios such as a thermal electron population or a structured jet to explain the data. Our observations demonstrate that millimeter light curves can exhibit some of the rich features more commonly seen in optical and X-ray afterglow light curves, motivating further millimeter wavelength studies of GRB afterglows.
\end{abstract}

\keywords{gamma-ray burst: general – gamma-ray burst: individual (GRB 210702A)}

\section{Introduction} \label{sec:intro}

The highly relativistic jets in long gamma-ray bursts (GRBs) are thought to be ejected by a black hole or a magnetar central engine that is formed following the catastrophic destruction of a massive Wolf-Rayet star due to their association with Type Ic supernovae \citep{Woosley2006}. In the fireball model for GRBs, the GRB itself is produced via emission from internal dissipation processes within the expanding jet, while the longer-lived broadband afterglow emission is produced within an external forward shock created by the expanding jet as it sweeps up the surrounding medium \citep{Meszaros1997,Sari1998,Piran1999,Piran2004,Kumar2015}. 

The forward shock model for GRB afterglows has been successful in explaining large swathes of observational data. In this basic picture, electrons are accelerated within the decelerating forward shock, which results in a power-law in energies that gives rise to a multi-segment broken power-law synchrotron emission spectrum characterised by a number of break frequencies whose evolution is dictated by the dynamics of the blast wave \citep{Meszaros1997,Sari1998}. Optical and X-ray observations have shown widespread compatibility with the synchrotron forward shock model \citep{Wang2015}, though extensions from the simplest case have been required to account for less common features such as achromatic steepenings, flares and plateaus \citep{Rhoads1999,Sari1999,Zhang2006}. 

Compared to X-ray and optical observations, radio follow-up of GRBs has been more sparse, despite the key role that radio observations can play in elucidating the physics of GRB jets \citep{Frail1997,Waxman1998,Berger2003,Chandra2012,Granot2014}. Radio observations uniquely sample the peak of the synchrotron spectrum and the synchrotron self-absorption break, allowing for the GRB jet energy, external density, and microphysical parameters to be determined via broadband modelling \citep{BarniolDuran2014,Laskar2015,Aksulu2022,Duncan2023}. At later times once the blast wave has transitioned to a non-relativistic, spherical state, radio observations may be used to perform calorimetry and thereby constrain the jet energetics \citep{Frail2000,Berger2004,Frail2005,Vanderhorst2008}. Very Long Baseline Interferometry observations have been used to constrain the source size and proper motion, for instance in GBR 030329 \citep{Taylor2005,Pihlstrom2007,Mesler2012}. Perhaps the most valuable contribution from studies at radio frequencies has been the detection of a second emission component due to reverse shock emission from a shock wave travelling back into the jet \citep{Kulkarni1999,Frail2000a,Laskar2013,Perley2014,Vanderhorst2014}, which has resulted in constraints on the jet's magnetization and initial bulk Lorentz factor \citep{Sari1999b,Kobayashi2003,Zhang2005,Laskar2018a,Laskar2018b}. 

Observationally, radio light curves often rise at early times and peak at a few days post-burst before declining \citep[at 8.5~GHz;][]{Chandra2012}, which is in contrast to X-ray and optical light curves which are generally decaying at all times. Lower frequency radio light curves may even peak on longer timescales of months to years \citep[e.g. GRB 030329;][]{Vanderhorst2008}. These peaks are commonly interpreted as the passage of the synchrotron peak frequency through the observing band, with lower frequencies peaking later than higher frequencies. However, sample studies have found that a large number of radio afterglows are challenging to interpret within standard forward shock models and may require non-standard scenarios such as a long-lived reverse shock, altered electron population, or a structured outflow to explain the data \citep{Panaitescu2004,Kangas2020,Levine2023,Laskar2023}. Radio light curves show a diversity of behaviour, with some showing smooth light curves and others showing flares \citep{Kulkarni1999,Soderberg2007,Schroeder2023}. 
At early times, diffractive and refractive interstellar scintillation can severely affect centimeter-band observations, hampering physical interpretation \citep[e.g. GRB 161219B;][]{Laskar2018b,Alexander2019}. In contrast, millimeter observations are usually not subject to scintillation and therefore any observed light curve feature must be intrinsic to the source. Despite its value, millimeter follow-up efforts have been more limited compared to the centimeter bands, particularly in the pre-ALMA era when existing observatories had limited sensitivity \citep{Postigo2012}.

Here we present X-ray, optical, millimeter and radio observations of the long GRB 210702A spanning 8 orders of magnitude in frequency (1 keV to 700 MHz) and almost 5 orders of magnitude in time (0.001 to 86 days). Our ALMA 97.5~GHz ($\approx3$~mm) light curve shows an unpredecedented rebrightening beginning at 8.2 days and increasing by a factor of $\approx2$ in brightness to reach a peak at 18.1 days before declining again. We find that a standard forward shock model can explain most of our multi-wavelength data prior to the millimeter rebrightening, but cannot explain the low frequency radio data and early X-ray light curve. We investigate a number of common explanations for rebrightenings in GRB afterglows and find that energy injection or a reverse shock from a late-time shell collision are plausible explanations, although we are unable to conclusively favour one scenario. 

We report all uncertainties at the $1\sigma$ level unless stated otherwise, and all magnitudes are in the AB system. We adopt a $\Lambda$CDM cosmology with $\Omega_m=0.315$ and $H_0=67.4$~km~s$^{-1}$~Mpc$^{-1}$ \citep{Planck2020}.

\section{Observations}
\subsection{Prompt emission}
The \Swift{} Burst Alert Telescope \citep[BAT;][]{BAT} was triggered by GRB 210702A at 19:07:13 UT on 2021 July 2 \citep{30351}. Refined BAT analysis indicated that the $\gamma$-ray mask-weighted light curve showed a single-peaked fast rise exponential decay-like structure with a measured duration of $T_{90}=138.2\pm47.6$~s. GRB 210702A was also detected by the CALET, AGILE, and Konus-Wind $\gamma$-ray detectors \citep{30362,30363,30366}. The Konus Wind fluence in the 20~keV -- 20~MeV range was $(2.5\pm0.2)\times10^{-4}$~erg cm$^{-2}$. The burst redshift of $z=1.160$ \citep{30357} corresponds to a luminosity distance of $D_L=2.53\times10^{28}$~cm, from which we calculate a burst isotropic $\gamma$-ray energy of $E_{\gamma,\mathrm{iso}}=(9.3\pm0.7)\times10^{53}$~erg.
We take the BAT trigger time as $T_0$ and reference all other times with respect to this time.

\subsection{X-ray}\label{sec:XRT}
The X-Ray Telescope aboard \Swift{} \citep[XRT;][]{XRT} began observing the field of the GRB 95.5~s after the BAT trigger and identified a bright new X-ray source consistent with the BAT position \citep{30351}. The first 310~s of data were taken in windowed timing (WT) mode while BAT was also capturing data, after which \Swift{} had to slew away. Observations began again in photon counting (PC) mode at 3561~s post-trigger. We downloaded the X-ray count-rate light curve and time-averaged WT and PC spectra from the online \Swift{}/XRT GRB Catalogue\footnote{\href{https://www.swift.ac.uk/xrt_live_cat/01058804/}{https://www.swift.ac.uk/xrt\_live\_cat/01058804/}}. We fitted the WT and PC mode spectra in Xspec version 12.13.0 with a photoelectrically absorbed power-law model (\verb|tbabs*ztbabs*pow|) taking into account Galactic and host galaxy absorption. The Galactic column density was fixed at $N_\mathrm{H,Gal}=1.19\times10^{21}$~cm$^{-2}$ \citep{Willingale2013} and the burst redshift at $z=1.160$. For the PC mode spectrum we derived a host galaxy column density of $N_\mathrm{H,host}=(2.3\pm0.6)\times10^{21}$~cm$^{-2}$ with a C-stat of 522 for 561 degrees of freedom. We obtained photon indices of $\Gamma=1.54\pm0.01$ and $\Gamma=1.95\pm0.03$ for the WT and PC mode spectra, respectively. The resulting spectral indices\footnote{Using $\beta_\mathrm{X}\equiv1-\Gamma$.} are therefore $\beta_\mathrm{X,WT}=-0.54\pm0.01$ and $\beta_\mathrm{X,PC}=-0.95\pm0.03$. Using the photon indices from our fits along with respective unabsorbed counts-to-flux conversion factors of $4.86\times10^{-11}$~erg~cm$^{-2}$~count$^{-1}$ and $4.54\times10^{-11}$~erg~cm$^{-2}$~count$^{-1}$ from our fits, we converted the 0.3--10~keV count-rate light curve into a 1~keV flux density light curve. 

\subsection{UV/Optical}
The \Swift{} Ultra-Violet Optical Telescope \citep[UVOT;][]{UVOT} started observing the BAT error region 104~s after the BAT trigger in the \textit{white} filter. A bright optical source consistent with the XRT position was identified at coordinates $\alpha=11^\mathrm{h}14^\mathrm{m}18.70^\mathrm{s}$, $\delta = -36^\circ 44^{\prime}50.0^{\prime\prime}$ (J2000) with an uncertainty of $0.42^{\prime\prime}$ \citep{30356}. The first 147~s \textit{white} filter exposure was saturated, along with a 44 s $u$-band exposure that began at 316~s post-trigger. UVOT recommenced observations just under one hour post-trigger, detecting the afterglow with a brightness of $b=15.15\pm0.03$~mag. Observations continued intermittently in all filters until 8.08~days post-trigger. We obtained the UVOT data from the \Swift{} Archive Download Portal\footnote{\href{https://www.swift.ac.uk/swift_portal/}{https://www.swift.ac.uk/swift\_portal/}} hosted on the UK \Swift{} Science Data Centre website, and extracted magnitudes using the \verb|uvotproduct| tool from the HEASoft\footnote{\href{https://heasarc.gsfc.nasa.gov/docs/software/lheasoft/}{https://heasarc.gsfc.nasa.gov/docs/software/lheasoft/}} \Swift{} FTOOLS software package, version 6.31.1. We employed a $5^{\prime\prime}$ aperture for the source region and a $10^{\prime\prime}$ aperture for the background region. For the saturated $u$-band exposure we adopt the AB magnitude measurement of $u=12.34\pm0.18$ mag calculated by \citet{Zhou2023} using their method for measuring photometries of moderately saturated UVOT sources. We do not include the $white$ filter exposures since the $white$ filter is too broad for our purposes. 

The MeerLICHT optical telescope \citep{Bloemen2016} was automatically triggered by the \Swift{} detection of GRB 210702A and began observing the BAT errorbox at 19:22:54 UT, just under 16 minutes after the BAT trigger time. Observations consisted of 60~s exposures in six optical filters following the filter sequence $quqgqrqiqz$, where the $q$ band is a wide filter spanning 440--720~nm and the $ugriz$ bands are standard SDSS filters. The optical afterglow was detected with a brightness of $q=12.84\pm0.02$~mag in our first exposure \citep{30354}. A total of 15 exposures were obtained over 25 minutes of observing before strong winds forced the telescope to close. We used the MeerLICHT pipeline (Vreeswijk et al., in prep.) to reduce the data, implementing standard CCD reduction tasks including astrometry and PSF photometry. The afterglow was well detected in all images.  

We additionally make use of optical photometry made public through Gamma-Ray Coordinates Network (GCN) Circulars: the X-Shooter acquisition camera on the Very Large Telescope (VLT) detected the afterglow in a 20~s exposure at 0.18~days post-trigger with a brightness of $r^{\prime}=16.91\pm0.01$~mag \citep{30357}, and the Chilescope RC-1000 telescope detected the afterglow at 0.19~days with a brightness of $r^{\prime}=17.01\pm0.05$~mag \citep{30365}.

\subsection{Radio}
\subsubsection{ALMA}
We obtained nine epochs of observations with the Atacama Large Millimeter/submillimeter Array (ALMA) in Band 3 (97.5~GHz) and three epochs of observations in Band 7 (343.5~GHz), through programs 2019.1.01484.T (PI Laskar) and 2019.1.01032.T (PI Perley). The Band 3 and Band 7 observations both employed two 4~GHz base-bands centred on 91.5 and 103.5~GHz, and 337.5 and 349.5~GHz, respectively. We downloaded the pipeline-generated images from the ALMA archive and measured the source flux using the Common Astronomy Software Application \citep[CASA;][]{McMullin2007} \verb|imfit| task.

\subsubsection{ATCA}\label{subsec:ATCA}
We observed GRB 210702A at seven epochs with the Australia Telescope Compact Array (ATCA) beginning at 3.50 days post-trigger. The first epoch consisted of observations in the 16.7, 21.2, 33 and 35 GHz bands, while all subsequent epochs consisted of observations at these same frequencies along with the 5.5 and 9 GHz bands. We used PKS 1934$-$638 as the flux and bandpass calibrator and PKS 1144$-$379 as the complex gain calibrator. At 33 and 35 GHz the flux calibrator was significantly fainter than the gain calibrator, so in these two bands we used the gain calibrator as our bandpass calibrator (after first deriving its intrinsic spectral index using PKS 1934$-$638) in order to derive an accurate bandpass solution. 

We used the \verb|MIRIAD| software package\footnote{\href{https://www.atnf.csiro.au/computing/software/miriad/}{https://www.atnf.csiro.au/computing/software/miriad/}} \citep{Miriad} for calibration and the CASA \verb|tclean| task for imaging the calibrated target visibilities. We combined the 33 and 35 GHz measurement sets prior to imaging in order to obtain a single 34 GHz flux measurement. At 5.5 GHz we combined the target visibilities from all epochs and imaged the resulting data set in order to create a sky model for the other sources in the field. We then performed imaging on the UV-subtracted target visibilities in order to measure the source flux. During epochs four through six the Array was in a configuration in which antenna CA06 was located far away compared to the other antennas. We excluded antenna CA06 while imaging the fifth epoch data as it produced images with a lower noise level. The epoch three and seven high frequency (16.7--35 GHz) gain calibrator
phases varied wildly as a result of poor weather. We therefore exclude these data from subsequent analysis. 

We measured the flux of the afterglow using two methods: using the CASA \verb|imfit| task on the target images; and UV-fitting the calibrated measurement sets with a point source model with the CASA \verb|uvmodelfit| task. The fluxes measured with both methods were generally found to agree within errors. We employ the UV fitting fluxes henceforth.

Additionally, \citet{Anderson2023} obtained early ($t<1$ day) radio observations of GRB 210702A with ATCA and found rapid variability in the radio light curves which they attribute to interstellar scintillation. Since short time-scale variability is not the focus of this paper, we take the average of the $3\sigma$ fluxes per observing band from their Table A1 which results in a single flux measurement in each of the 5.5, 9, 16.7 and 21.2 GHz bands at $\approx0.53$ days. 

\subsubsection{MeerKAT}
We observed the radio afterglow with the MeerKAT radio telescope in the L band (1.28 GHz) at five epochs through observing program SCI-20210212-TL-01 (PI Laskar). We used J0408$-$6545 as the flux and bandpass calibrator for the first and last epochs and J1939$-$6342 as the flux and bandpass calibrator for the intervening epochs, while J1154$-$3505 was used as the complex gain calibrator for all epochs. The total integration time on source was 0.65~hr per epoch. We flagged, calibrated and imaged the data using the \verb|oxkat| pipeline\footnote{\href{https://github.com/IanHeywood/oxkat}{https://github.com/IanHeywood/oxkat}} \citep{Heywood2020}, and measured the flux of the afterglow with the \verb|imfit| task in CASA.  

\subsubsection{GMRT}
We obtained five epochs of observations with the Giant Metrewave Radio Telescope (GMRT) through program 40\_084 (PI Laskar) beginning at 8.6 days post-trigger. Observations were carried out in Band 4 (700~MHz) with a 400~MHz bandwidth, employing 3C286 as the flux and bandpass calibrator, and J1154$-$350 as the complex gain calibrator. We flagged and calibrated the data using standard reduction techniques in CASA, and thereafter used the Inter-University Institute
for Data Intensive Astronomy (IDIA) \verb|processMeerKAT| software pipeline\footnote{\href{https://idia-pipelines.github.io/docs/processMeerKAT}{https://idia-pipelines.github.io/docs/processMeerKAT}} (Collier, Sekhar, Frank, Taylor, et al., in prep.) to perform imaging and two rounds of phase self-calibration followed by one round of amplitude and phase self-calibration. No radio source was visible at the GRB position in the images from the first four epochs, so we report upper limits as three times the RMS noise at the afterglow position, calculated using the CASA task \verb|imstat|. We detected the afterglow in our fifth and final epoch image and measured the flux density using \verb|imfit|. 

We report all X-ray, optical and radio flux measurements in Table \ref{tab:flux}.

\section{Multi-wavelength modelling}\label{sec:results}
We interpret our observations within the synchrotron forward shock model of GRB afterglows. In this model, a collimated relativistic blast wave sweeps up the surrounding circumburst medium forming a shock front in which electrons are accelerated into a power-law distribution in energies characterised by a spectral index $p$. The electron distribution gives rise to a multi-segment broken power-law emission spectrum consisting of spectral breaks at three frequencies: $\nu_m$ is the frequency of the spectral break corresponding to electrons with the minimum energy in the distribution; $\nu_c$ is that of the break above which electrons are undergoing cooling; and $\nu_a$ is the frequency of the break below which synchrotron radiation is self-absorbed. The evolution of the spectral breaks is dictated by the dynamics of the blast wave, for which we assume the \citet{BM1976} solution for a relativistic explosion. The synchrotron forward shock model is described in detail in \citet{Sari1998} and \citet{Granot2002}. We follow the convention with $F_{\nu}\propto t^{\alpha}\nu^{\beta}$ henceforth.

\subsection{The millimeter rebrightening}

\begin{figure}
\includegraphics[width=\columnwidth]{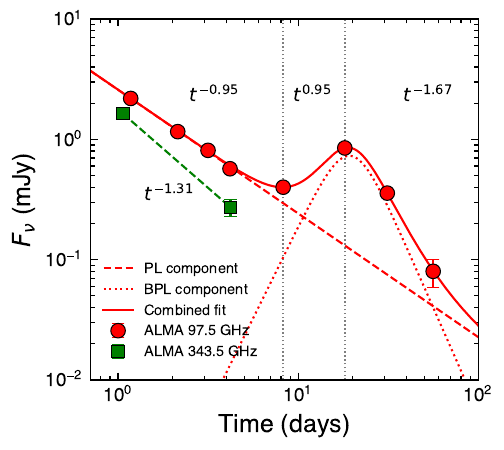}
\caption{ALMA 97.5~GHz and 343.5~GHz light curves for GRB 210702A. The ALMA 97.5~GHz light curve shows a clear rebrightening peaking at $\approx18$~days. We show the combined PL+BPL fit to the 97.5~GHz light curve as well as each model component. The results of our fit are presented in Table \ref{tab:LC_fit}. The decay rates from direct fits to each light curve segment are indicated above (below) the 97.5~GHz (343.5~GHz) light curve.}
\label{fig:ALMA_LC}
\end{figure}

\begin{figure}
\includegraphics[width=\columnwidth]{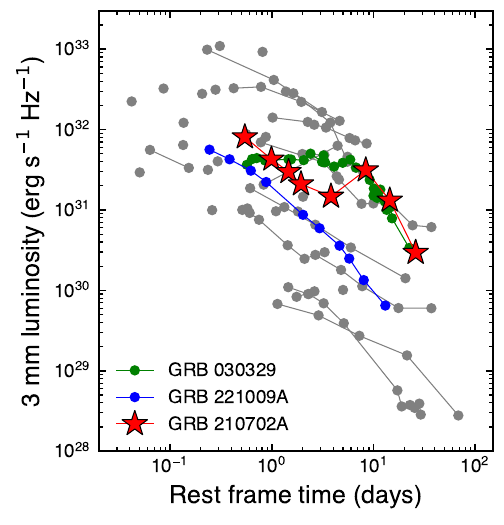}
\caption{Spectral luminosity at a wavelength of 3~mm for GRB 210702A and a sample of long GRBs (gray) from \citet{Eftekhari2022}. We highlight GRB 030329 and GRB 221009A, two nearby ($z=0.1685$ and $z=0.151$) and extremely bright bursts. GRB 210702A is the first GRB with a clear rebrightening at millimeter wavelengths, though its luminosity is at all times by no means exceptional.}
\label{fig:3mm_comparison}
\end{figure}

The most striking feature of our multi-wavelength data set is the ALMA 97.5~GHz light curve, as shown in Figure \ref{fig:ALMA_LC}. After five epochs of clear fading behaviour the light curve begins to rebrighten at 8.2 days post-trigger, rising to a peak at 18.1 days before declining again. This rebrightening by a factor of $\approx2$ in flux is, to our knowledge, the first such rebrightening seen in a millimeter afterglow light curve. Although the light curve behaviour is unusual, the 3 millimeter luminosity of GRB 210702A is typical when compared to other long GRB millimeter light curves (Figure \ref{fig:3mm_comparison}).

To characterise this light curve further, we fit a model comprising the sum of a power-law (PL) and a smoothly broken power-law (BPL), where the BPL follows the functional form
\begin{equation}\label{eq:BPL}
    F(t)=F_0 \left[\left(\frac{t}{t_b}\right)^{-\alpha_1\omega}+\left(\frac{t}{t_b}\right)^{-\alpha_2\omega}\right]^{-1/\omega},
\end{equation}
with $F_0$ as a normalising flux, $\alpha_1$ and $\alpha_2$ as the pre- and post-break temporal indices, $t_b$ as the break time, and $\omega$ as a smoothness parameter which sets the sharpness of the break. Larger values of $\omega$ ($>1$) correspond to a sharper break in the light curve. The results of our light curve fit in Table \ref{tab:LC_fit}  show that the power-law component decays with $\alpha_\mathrm{97.5~GHz}=-1.03\pm0.02$. For the broken power-law component we find that a smoother break with $\omega=1$ results in a better fit than sharper breaks. The rising and decaying indices for this component are $\alpha_1=3.04\pm0.78$ and $\alpha_2=-3.41\pm1.60$ with a break time of $t_b=19.5\pm4.3$ days. On the other hand, direct fits to the rising and decaying segments (Figure \ref{fig:ALMA_LC}) result in $\alpha_\mathrm{97.5GHz,rise}=0.95$ and $\alpha_\mathrm{97.5GHz,decay}=-1.67\pm0.17$, respectively, while the 343.5~GHz light curve decays with $\alpha=-1.31$. We return to this in our investigation of the millimeter rebrightening in Section \ref{sec:mm_rebrightening}.

\begin{deluxetable}{lc}
\tablecaption{ALMA 97.5~GHz light curve fit parameters\label{tab:LC_fit}}
\tablehead{\colhead{Parameter} & \colhead{Value}}
\startdata
PL $\alpha$ & $-1.03\pm0.02$\\
BPL $\alpha_1$ & $3.04\pm0.78$\\
BPL $\alpha_2$ & $-3.41\pm1.60$\\
BPL $t_b$ (days) & $19.50\pm4.33$\\
\enddata
\end{deluxetable}

\subsection{X-ray and UV/optical temporal evolution}\label{subsec:temporal}

\begin{figure*}
\gridline{\fig{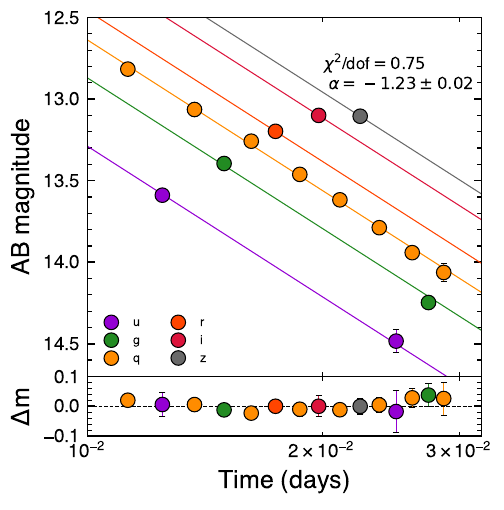}{0.45\textwidth}{(a)}
          \fig{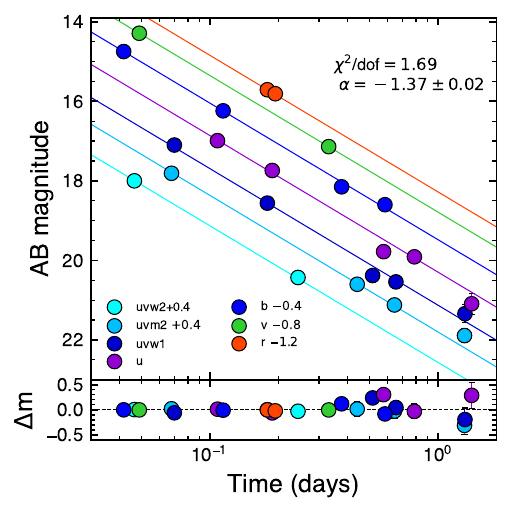}{0.45\textwidth}{(b)}}
\caption{Power-law light curve fits to the MeerLICHT photometry (a) and \Swift{}/UVOT photometry including the X-Shooter and Chilescope $r$-band measurements (b). The power-law decay rate was constrained to be the same across each band during fitting. The MeerLICHT residuals show evidence for a steepening in the light curve at $\approx0.016$ days.}
\label{fig:UV_opt_LCs}
\end{figure*}

The X-ray and UV/optical band data show declining light curves throughout our observations. Fitting the MeerLICHT and \Swift{} UV/optical data separately with power-laws in time, we find that the optical temporal decay rate during the MeerLICHT observations is $\alpha=-1.23\pm0.02$ while the decay rate during the \Swift{} UV/optical observations is $\alpha=-1.37\pm0.02$ (see Figure \ref{fig:UV_opt_LCs}), indicating a steepening in the optical decay rate between the two sets of observations. The residuals from the power-law fit to the MeerLICHT observations show evidence for the steepening occurring at $\approx0.016$ days. 

\begin{figure}
\centering
\includegraphics[width=\columnwidth]{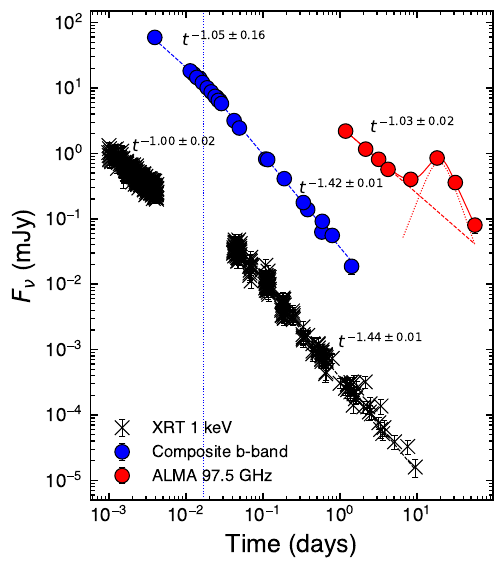}
\caption{X-ray, composite $b$-band and ALMA 97.5~GHz light curves. We show the power-law fits to the WT-mode and PC-mode X-ray light curve segments (black dashed lines), and the broken power-law fit to the optical light curve (blue dashed line). The vertical dotted line denotes the break time of the optical light curve fit, $t_b=0.017\pm0.002$~days. We show the light curve fit from Figure \ref{fig:ALMA_LC}.  }
\label{fig:LC_fits}
\end{figure}

We measure the $u$- to $v$-band spectral index during each set of observations by interpolating our light curve fits in Figure \ref{fig:UV_opt_LCs} to a common time, and find that the MeerLICHT in-band spectral index of $\beta_\mathrm{MeerLICHT}=-1.42\pm0.15$ is consistent with the UVOT in-band spectral index of $\beta_\mathrm{UVOT}=-1.46\pm0.11$. To characterise the optical temporal evolution further, we create a composite $b$-band light curve by transforming the UVOT $u$, $b$ and $v$ bands along with the MeerLICHT $u$, $g$ and $q$ bands to a common observing frequency equivalent to the UVOT $b$-band using a spectral index of $\beta=-1.44$, the average spectral index from our spectral fits to the MeerLICHT and UVOT data. Since the optical light curve steepens between the MeerLICHT and UVOT observations, we fit the light curve with a smoothly broken power-law following Equation \ref{eq:BPL}. The best-fit light curve (see Figure \ref{fig:LC_fits}) has pre- and post-break temporal indices of $\alpha_1=-1.05\pm0.16$ and $\alpha_2=-1.42\pm0.01$, respectively, favouring a sharper break ($\omega\approx29$) with a break time of $t_b=0.017\pm0.002$ days which is consistent with the evidence for a break seen in our residuals from the power-law fit to the MeerLICHT data. The steepening of the light curve to $\alpha=-1.42\pm0.01$ is unlikely to be due to a jet break since post-jet break decay is predicted to decay as $t^{-p}$ with $p$ between 2 and 3 \citep{Sari1999}. If lateral spreading of the jet is not significant, the predicted steepening is $\Delta\alpha=-0.75$ in an interstellar medium-like (ISM) circumburst medium or $\Delta\alpha=-0.5$ in a stellar-wind medium \citep{Granot2002}. Both of these cases are still steeper than the measured $\Delta\alpha=-0.37\pm0.16$, though the wind case does agree within the errors. If the temporal steepening is due to the passage of the cooling break we would expect a change of $\Delta\alpha=-0.25$ which is shallower than the measured $\Delta\alpha=-0.37\pm0.16$, but also consistent within the errors. We return to this in Section \ref{subsec:wind}. 

The X-ray light curve consists of two segments separated by a period during which \Swift{} had to slew away. Fitting each segment with a simple power-law (see Figure \ref{fig:LC_fits}), we find that the early WT-mode decay rate of $\alpha_\mathrm{WT}=-1.00\pm0.02$ is consistent with the early optical evolution, while the later PC-mode decay rate of $\alpha_\mathrm{PC}=-1.44\pm0.01$ is in very close agreement with the post-break optical decay rate of $\alpha_2=-1.42\pm0.01$. Although the pre- and post-break X-ray and optical decay rates are very similar and may suggest achromatic X-ray to optical evolution, we note that the X-ray to optical $b$-band spectral index was $\beta_\mathrm{O-X}=-0.91$ at 0.004 days and $\beta_\mathrm{O-X}=-0.77$ at 0.1 days, therefore ruling out achromatic evolution between the two bands. We return to this in Section \ref{subsec:wind} below.

\subsection{Stellar wind medium}\label{subsec:wind}
We now investigate whether a standard forward shock model can explain the observed spectral and temporal characteristics of our X-ray, UV/optical and early millimeter data. From 0.2--10 days our X-ray and UV/optical light curves decline as power-laws with similar indices of $\alpha~\approx-1.43$, suggestive of a `normal' afterglow decay and that both bands may lie within the same spectral regime. During slow-cooling ($\nu_m<\nu_c$) there are two spectral regimes to consider for our declining X-ray and optical light curves: either $\nu_m<\nu_c<\nu_\mathrm{O}<\nu_\mathrm{X}$ or $\nu_m<\nu_\mathrm{O}<\nu_\mathrm{X}<\nu_c$. 

In the regime with $\nu_m<\nu_c<\nu_\mathrm{O}<\nu_\mathrm{X}$ we expect a spectral index of $\beta=-p/2$ and a temporal decay rate of $\alpha=(2-3p)/4$, regardless of the circumburst medium. Using the average of the observed optical and X-ray temporal indices of $\alpha=-1.43\pm0.01$ results in $p=2.57\pm0.01$ and a spectral index of $\beta=-1.28\pm0.01$. Correcting our optical SED derived from the \Swift{}/UVOT light curve for a Galactic extinction of $A_V=0.3271$~mag along the line of sight \citep{SF2011}, the $b$- to $r$-band spectral index of $\beta_\mathrm{O}=-0.63\pm0.08$ is clearly in disagreement with the predicted spectral index of $\beta=-1.28\pm0.01$, ruling out this scenario.

In the regime with $\nu_m<\nu_\mathrm{O}<\nu_\mathrm{X}<\nu_c$ we expect a spectral index of $\beta=(1-p)/2$ and temporal indices of $\alpha_\mathrm{ISM}=3(1-p)/4$ for an ISM-like circumburst medium or $\alpha_\mathrm{wind}=(1-3p)/4$ for a stellar wind medium. For the ISM case we derive $p=2.91\pm0.01$ and a spectral index of $\beta=-0.96\pm0.01$ from our temporal index of $\alpha=-1.43\pm0.01$. This was the scenario favoured by \citet{Anderson2023}. Although this spectrum agrees with the PC-mode X-ray spectral index of $\beta_\mathrm{X}=-0.95\pm0.03$ (Section \ref{sec:XRT}), it is steeper than the optical index of $\beta_\mathrm{O}=-0.63\pm0.08$ and the optical-to-X-ray index of $\beta_\mathrm{O-X}=-0.77$ calculated at 0.1 days. For the stellar wind case we derive $p=2.24\pm0.01$ from our observed light curve evolution and an expected spectral index of $\beta=-0.62\pm0.01$, which is in close agreement with the optical spectral index of $\beta_\mathrm{O}=-0.63\pm0.08$ and therefore our preferred scenario. With $p=2.24\pm0.01$ the predicted spectral index above the cooling break is $\beta=-1.12\pm0.01$, which is slightly steeper than our X-ray spectral index of $\beta_\mathrm{X}=-0.95\pm0.03$. In a stellar wind medium, the frequency of the cooling break, $\nu_c$, rises as $t^{1/2}$. A smooth cooling break \citep{Uhm2014} with $\nu_c$ close to the X-ray band is a possible explanation for the similar temporal indices of the X-ray and optical bands, but steeper X-ray spectrum compared to the optical.

Further support for a stellar wind medium is found when considering the early millimeter data. At the time of the first ALMA 97.5~GHz detection at 1.17 days, we measure a 97.5~GHz to optical $r$-band spectral index of $\beta_\mathrm{97.5GHz-O}=-0.47$, which is shallower than the optical in-band spectral index $\beta_\mathrm{O}=-0.63\pm0.08$ and optical-to-X-ray index of $\beta_\mathrm{O-X}=-0.77$. Moreover, extrapolating our 97.5~GHz light curve fit to the time of the first 343.5~GHz detection at 1.06 days yields a 97.5~GHz to 343.5~GHz spectral index of $\beta_\mathrm{97.5-343.5GHz}=-0.31$, indicating that the spectrum is flattening off towards lower frequencies and that $\nu_m$ lies below the millimeter bands. In an ISM medium the peak flux of the synchrotron spectrum remains constant with time, whereas in a stellar wind medium the peak flux declines as $t^{-1/2}$. Assuming that $\nu_m$ is within the 97.5~GHz band at the time of our first detection at 1.17 days, the corresponding peak flux is $F_{\nu_m}\approx2$~mJy. Our $u$-band detection at 0.004 days has a flux density of $F_u\approx42$~mJy, which is more than an order of magnitude larger than our constraint on the peak flux from our millimeter data at 1.17 days and therefore clearly rules out an ISM medium. 

The peak frequency for both an ISM and wind medium evolves as $t^{-3/2}$. If we assume a stellar wind medium in which $\nu_m$ is close to the 97.5~GHz band with a flux of $F_{\nu_m}\approx2$ mJy at 1.17 days, we expect the peak flux to be $F_{\nu_m}\approx34$~mJy at 0.004 days and $\nu_m$ to be at a frequency $\nu_m\approx4.9\times10^{14}$~Hz, which corresponds to the optical $r$-band. Since the $u$ band must be above $\nu_m$ due to the declining optical light curve and negative spectral slope observed in the MeerLICHT data, $\nu_m$ in the $r$-band at 0.004 days is consistent with our observations. The expected peak flux of $F_{\nu_m}\approx34$~mJy is only a factor 1.2 smaller than our $u$-band detection of $F_u\approx42$~mJy at this time, and so we still regard a stellar wind medium as being consistent with our observations. 

The UV/optical and millimeter light curves initially have temporal decay rates with $\alpha\approx-1$ compared to the later decay rate of $\alpha\approx-1.43$ observed in the optical and X-rays. This shallower decay might be due to the proximity of $\nu_m$ to the observing band, causing the light curve to decay less steeply initially before transitioning to the expected decay rate. The shallow decay with $\alpha=-1.00\pm0.02$ observed in the X-rays prior to 0.01 days, however, cannot be explained via this interpretation since $\nu_m$ should be far below the X-ray band at this time. An explanation for the shallower X-ray decline could be continuous energy injection into the GRB blast wave which results in an observed plateau or shallow decay, an effect which has been observed extensively in X-ray light curves \citep{Zhang2006,Nousek2006}.

\subsection{Radio evolution} \label{subsec:radio}
We have shown in the previous section that the optical, X-ray data after 0.01 days, and millimeter data prior to the millimeter wavelength rebrightening can be explained with a standard forward shock model in a stellar wind medium with $p\approx2.2$. We now consider the implications of our lower frequency radio data. 

\begin{figure}
\centering
\includegraphics[width=\columnwidth]{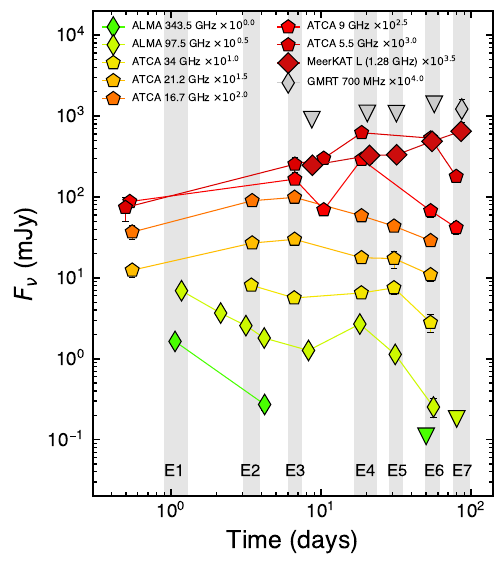}
\caption{Radio light curves separated by observing band. The vertical grey columns indicate the epochs for which we show the corresponding radio SEDs in Figure \ref{fig:Radio_SEDs}. We shift each light curve vertically in flux for clarity, while upper limits are shown as upside down triangles. We advise caution in interpreting the significant dip seen in the 9~GHz light curve at 10.4 days since this epoch of ATCA data was strongly affected by phase instabilities (see Section \ref{subsec:ATCA}). }
\label{fig:Radio_LC}
\end{figure}

\begin{figure*}
\centering
\includegraphics[width=\textwidth]{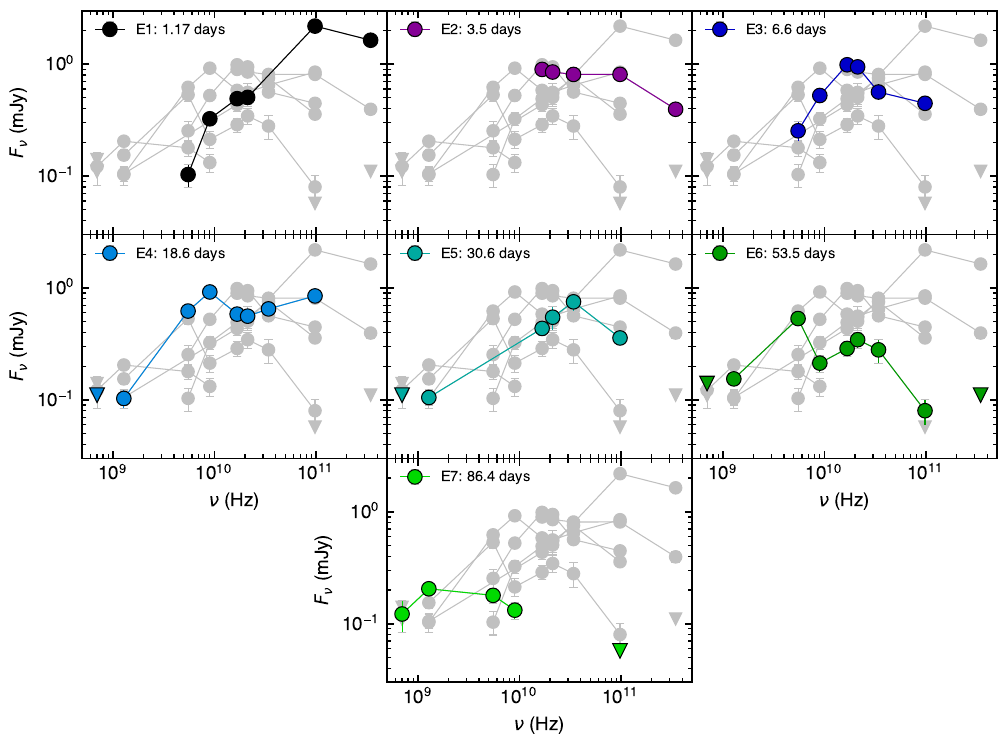}
\caption{SEDs at the epochs indicated in Figure \ref{fig:Radio_LC}. Epochs 1 through 7 correspond to approximate times of 1.17, 3.5, 6.6, 18.6, 30.6, 53.5 and 86.4 days post-trigger. For Epoch 1 we interpolate the ATCA fluxes to a time of 1.17 days, while for Epochs 2 and 3 we interpolate the 97.5~GHz and 343.5~GHz flux to their respective times. Upper limits are shown as upside down triangles. The rising segment in the Epoch 1 SED has a spectral slope of $\beta=0.93\pm0.06$.}
\label{fig:Radio_SEDs}
\end{figure*}

Figure \ref{fig:Radio_LC} shows our radio light curves separated by observing band and instrument. We have extracted SEDs at seven epochs indicated by the vertical grey regions, which are shown in Figure \ref{fig:Radio_SEDs}. The ALMA 97.5~GHz rebrightening begins at 8.2 days and rises to a peak at 18.1 days. Prior to the start of the rebrightening at 8.2 days we assume that the rebrightening does not have a significant effect on the light curves nor SEDs. The early ATCA observations obtained by \citet{Anderson2023} allow us to construct an SED at the time of our first ALMA observations by interpolating the light curves using their power-law behvaiour. The Epoch 1 SED shows a clearly rising spectrum from 5.5~GHz to 97.5~GHz measurement with a spectral index of $\beta_\mathrm{5.5-97.5GHz}=0.93\pm0.06$. Such a rising spectrum is suggestive of optically thick synchrotron emission, which we use to place a tentative constraint on the self-absorption break of $\nu_a\approx97.5$~GHz at 1.17 days. From our arguments in Section \ref{subsec:wind}, this places $\nu_a$ very close to $\nu_m$ at this time. In Section \ref{subsec:modelling} below we will demonstrate that no standard forward shock model can accommodate such a high self absorption frequency.   

Although $\beta_\mathrm{5.5-97.5GHz}=0.93\pm0.06$ is not as steep as the theoretically-predicted slope of $\nu^2$ for synchrotron self-absorbed emission, a smooth self-absorption break may account for the shallower measured slope. The rising ATCA light curves at $t<8$~days (see Figure \ref{fig:Radio_LC}) with $\alpha\approx1$ are also suggestive of optically thick emission, since the predicted temporal evolution below $\nu_a$ is $t^1$ in a stellar wind medium. By the time of our Epoch 2 SED at 3.5 days we see a flat spectrum from 16.7 to 97.5~GHz. This allows us to place a constraint on the self-absorption break of $\nu_a<16.7$~GHz, and that $\nu_a$ must have evolved faster than $t^{-1.6}$ between these two epochs. Such a fast evolution of the self-absorption break is inconsistent with any known evolution. The declining spectrum from 97.5 to 343.5~GHz appears to be consistent with $\nu_m$ having passed through the millimeter bands as demonstrated in Section \ref{subsec:wind} above. The Epoch 3 SED at 6.6 days provides a second clear case of a self-absorbed spectrum with a spectral index $\beta=1.03\pm0.16$ measured from 5.5 to 16.7~GHz, and a self-absorption break at $\nu_a\approx16.7$~GHz. The fact that the self-absorption break evolved as $t^{-1.6}$ between Epochs 1 and 2 and not at all between Epoch 2 and 3 demonstrates that the optically-thick radio data cannot be reconciled within any standard forward shock model. 

Throughout the millimeter rebrightening, our radio light curves do not show obviously achromatic behaviour. The low frequency 1.4, 5.5 and 9~GHz bands rise and even peak when the 97.5~GHz light curve is rising and peaks, whereas the higher frequency bands at 16.7 and 21.2~GHz do not show simultaneous rises. The 34~GHz band appears to rise slowly at the start of the 97.5~GHz rebrightening but peaks later at $\approx30.6$ days. We note that the omission of higher frequency data from our third epoch of ATCA data at 10.4 days (Section \ref{subsec:ATCA}) leads to a loss of possibly valuable temporal information. Following the 97.5~GHz peak, all of the radio light curves besides the 34 and 1.28~GHz light curves decline. Since the MeerKAT band is in the optically thick regime throughout our observations (see Figure \ref{fig:Radio_SEDs}), the rising L band light curve appears consistent with rising self-absorbed synchrotron emission \citep{Granot2002}. 

Before, during, and after the rebrightening, our radio SEDs show complex behaviour. Epoch 4 coincides in time with the 97.5~GHz peak and the fluxes measured in that epoch show tentative evidence for an additional spectral component peaking at 97.5~GHz as seen in the rising spectrum from 16.7 to 97.5 GHz. The Epoch 5 and 6 SEDs show similar peaks at 34~GHz and 21.2~GHz, respectively. However, our radio SEDs are characterised by a number of sharp jumps which may not be intrinsic to the source since we do not expect synchrotron radiation from a relativistic blast wave -- even with additional spectral components -- to result in such sharp jumps in the spectrum. It may be the case that the small scale features in our light curves and SEDs are a result of unaccounted-for systematic uncertainties from the calibration of our ATCA data or due to interstellar scintillation (Section \ref{subsec:discuss_FS_models}). We discuss our radio data taken during and after the rebrightening further in Section \ref{sec:mm_rebrightening}.

\section{Afterglow modelling with ScaleFit}\label{subsec:modelling}
Constraints on the three spectral break frequencies ($\nu_a$, $\nu_m$, $\nu_c$) and the peak flux of the synchrotron spectrum can be used to derive intrinsic blast wave parameters of the forward shock and may also be used to test whether a standard forward shock model is compatible with our data. In Section \ref{subsec:wind} we constrained the peak flux to $F_{\nu_m}\approx2$ mJy at 1.17 days, the time of our first 97.5~GHz observation. We also have $\nu_m$ close to the 97.5~GHz band at this time, so $\nu_m\approx90$~GHz. We expect the cooling break to cross the X-ray band during our PC-mode observations, so $\nu_c\approx2.4\times10^{17}$~Hz at 0.5 days. Our Epoch 1 radio SED places a constraint on the self-absorption frequency of $\nu_a\approx97.5$~GHz at 1.17 days. Solving the system of four equations describing the locations of the spectral breaks and their flux densities in a wind medium (see Table 2 in \citet{Granot2002}) results in an unphysical value for the fraction of shock internal energy partitioned to electrons of $\epsilon_e>1$, which is driven primarily by our constraint on the self-absorption break. We therefore do not expect a standard wind-medium forward shock model to be able to account for the observed optically thick radio emission prior to the 97.5~GHz rebrightening at 8.2 days. We note that similar incompatibilities with standard forward shock models resulting from radio observations have been observed in other GRB afterglows \citep{Marongiu2022,deWet2023a,Laskar2023,Schroeder2023}. We discuss this further in Section \ref{subsec:discuss_FS_models}.

We perform theoretical modelling using the \verb|ScaleFit| software package \citep{Ryan2015,Aksulu2020,Aksulu2022}, which is based on the \verb|BoxFit| set of high-resolution hydrodynamic simulations of GRB explosions \citep{VanEerten2012}. \verb|ScaleFit| generates afterglow spectra and light curves of forward shock emission in both ISM or stellar wind environments during the deceleration phase after reverse shock crossing, and can take into account the jet break effect as well as the non-relativistic transition. Non-standard effects such as reverse shock emission, energy injection, plateaus, or flares, however, are not included. \verb|ScaleFit| accepts 10 free parameters: the isotropic-equivalent energy of the blast wave $E_\mathrm{K,iso}$, circumburst density $n_0$\footnote{In a wind medium the $A_\star$ parameter is commonly used as a measure of the density, as defined in \citet{Chevalier2000}. The $n_0$ parameter is related to the $A_\star$ parameter via $n_0=A_\star\times29.89$~cm$^{-3}$, where $n_0$ is the number density at a reference distance of $10^{17}$~cm. }, opening angle of the jet $\theta_j$, electron energy distribution spectral index $p$, fraction of accelerated electrons within the shock $\xi_N$, fractions of shock internal energy distributed to electrons and magnetic fields $\epsilon_e$ and $\epsilon_B$, burst redshift $z$, luminosity distance $d_L$, and observer angle relative to the jet axis $\theta_\mathrm{obs}$. 

We fix the burst redshift to $z=1.160$ and the luminosity distance to the value computed using our adopted cosmology, $d_L=2.52\times10^{28}$~cm. We assume that the observer is looking directly down the axis of the jet ($\theta_\mathrm{obs}=0$) and that the fraction of shock-accelerated electrons is unity despite the fact that $\xi_N$ is degenerate with respect to the parameters $\{E_\mathrm{K,iso},n_0,\epsilon_e,\epsilon_B\}$, as shown by \citet{Eichler2005}. To account for Galactic dust extinction we correct our UV/optical data using the \citet{Fitzpatrick1999} Milky Way extinction law with $R_V=3.1$ and $A_V=0.3271$~mag. Furthermore, we adopt the Small Magellanic Cloud (SMC) extinction curve from \citet{Pei1992} to account for possible host galaxy extinction, which is a free parameter in our modelling. 

We conduct a Markov Chain Monte Carlo (MCMC) exploration of our \verb|ScaleFit| parameter space using the \verb|emcee| Python package \citep{Foreman2013} following the implementation described in detail in \citet{deWet2023b}. The Lyman limit of 912~\r{A} at the host redshift lies squarely within the $uvw2$ ultraviolet band so these flux measurements are likely to be severely affected by photoelectric absorption. We therefore exclude these data from our modelling, along with the radio observations following the start of the 97.5~GHz rebrightening at 8.2 days. We also exclude the early-time X-ray data ($t<0.01$~days), the first epoch of ATCA data ($t\approx0.5$ days), and the lower frequency ($<16.7$~GHz) radio data at 6.6 days since we do not expect our model to be able to account for these data following the arguments in Sections \ref{subsec:wind} and \ref{subsec:radio}. We assume a spherical jet ($\theta=\pi/2$) since our observations show no clear evidence for a jet break.  

\begin{figure*}
\centering
\includegraphics[]{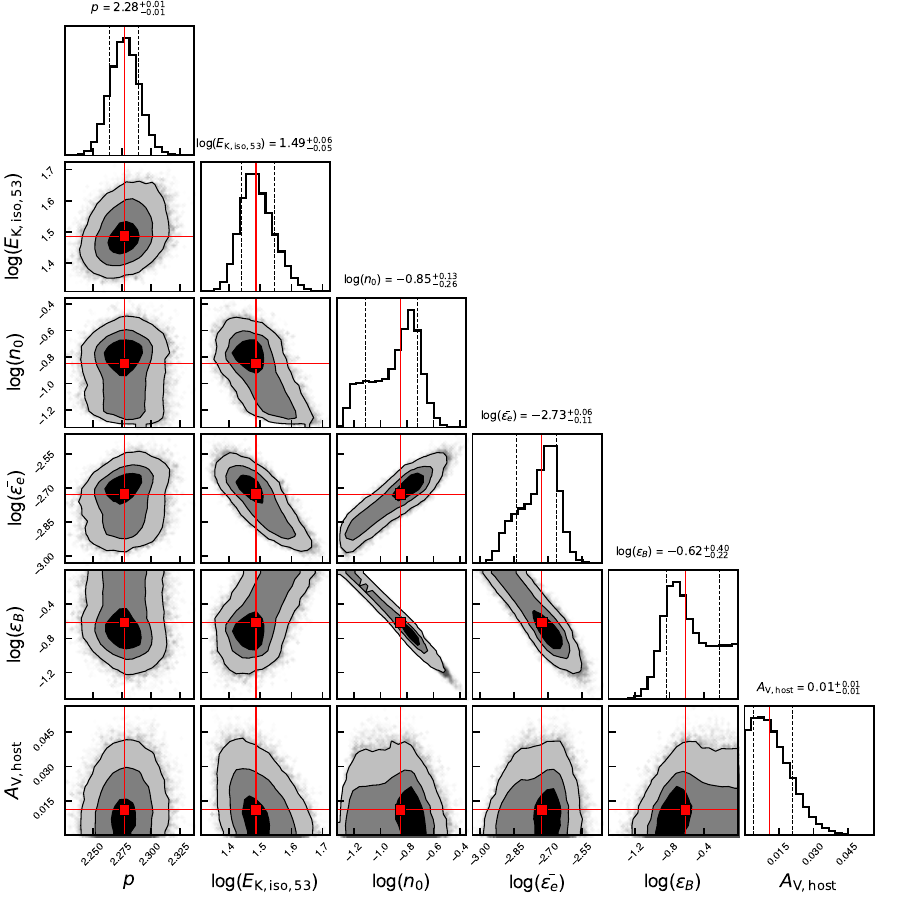}
\caption{Corner plot from our MCMC analysis. Contours denote the $1\sigma$, $2\sigma$ and $3\sigma$ levels. Red lines correspond to the median values in the marginalised distributions. }
\label{fig:MCMC}
\end{figure*}

\begin{figure*}
\centering
\includegraphics[width=\textwidth]{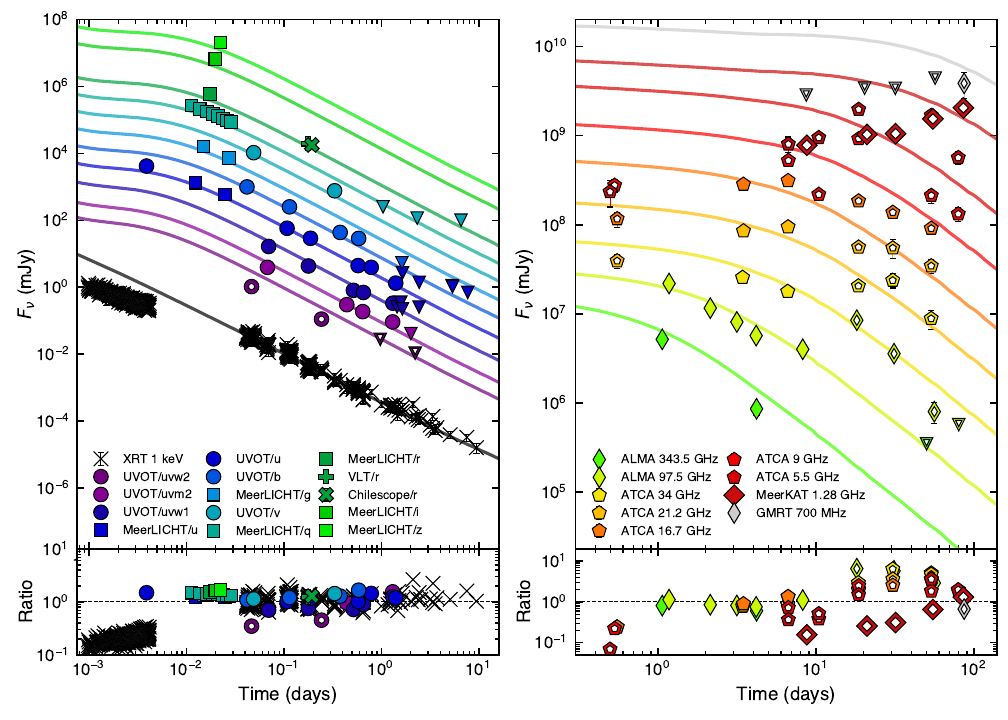}
\caption{Model light curves associated with the highest-likelihood forward shock model from our MCMC analysis. We separate the data by observing band and instrument, and shift each light curve by a factor of $10^{0.5 n}$ in flux, where $n$ is an integer. Open symbols denote those data which were not used to fit the model. Upside-down triangles indicate upper limits. The lower panels present the ratio of the measured flux with respect to the highest-likelihood model. The failure of the forward shock model to explain all of the data is discussed further in Section \ref{subsec:discuss_FS_models}.}
\label{fig:model_LCs}
\end{figure*}

\begin{figure}
\centering
\includegraphics[width=\columnwidth]{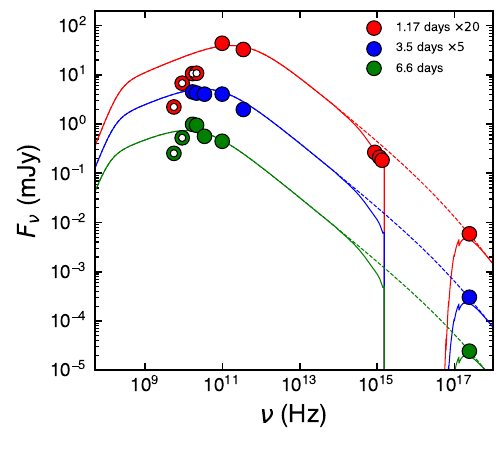}
\caption{Broadband SEDs at three epochs along with model SEDs corresponding to our highest-likelihood model. Dashed lines indicate the unabsorbed synchrotron spectrum, while solid lines take into account photoelectric absorption and Galactic and host galaxy extinction. The radio data correspond to the first three SEDs in Figure \ref{fig:Radio_SEDs}, respectively. Open symbols indicate those data which were not used to fit the model. }
\label{fig:model_SEDs}
\end{figure}

The results of our modelling are presented in Table \ref{tab:MCMC} and Figure \ref{fig:MCMC}, while the light curves corresponding to the highest-likelihood model are shown in Figure \ref{fig:model_LCs}. This model has $p=2.27$ which is close to the value of $p=2.24$ derived from our late-time optical and X-ray light curves in Section \ref{subsec:wind}. The spectrum is in fast cooling with $\nu_c<\nu_m$ until $\nu_m$ crosses $\nu_c$ at 0.0011 days at a frequency of $\nu\approx10^{16}$~Hz which is between the optical and X-ray bands. The cooling break $\nu_c$ begins crossing the X-ray band midway through our PC-mode observations which explains the steeper observed X-ray spectral index of $\beta_\mathrm{X}=-0.95\pm0.03$ compared to the optical index of $\beta_\mathrm{O}=-0.63\pm0.08$, while the similar temporal decay observed in both bands is a result of the smooth cooling break. By $\approx0.01$ days $\nu_m$ has passed through all of the optical bands, consistent with the observed steepening of the optical light curve from $\alpha_\mathrm{O}=-1.05$ to $\alpha_\mathrm{O}=-1.44$ at 0.017 days as seen in Figure \ref{fig:LC_fits}. At the time of our first ALMA observations we have $\nu_m\approx330$~GHz which is close to the 343.5~GHz ALMA band. The shallow decay observed in the early optical and millimeter light curves is a result of the proximity of $\nu_m$ to the observing band, consistent with our arguments in Section \ref{subsec:wind}. As we expected, the optically thick spectra observed in the Epoch 1 and 3 radio SEDs are incompatible with our highest-likelihood model, as shown in Figure \ref{fig:model_SEDs}. At 6.6 days the self-absorption frequency is at $\nu_a\approx110$~MHz, an order of magnitude below the observed break at 16.7 GHz. We return to this in Section \ref{subsec:discuss_FS_models}.

No jet break is seen in our data prior to the start of the millimeter rebrightening at 8.2 days, so we can place a lower limit on the opening angle using the inverted form of Equation 5 from \citet{Chevalier2000}:
\begin{equation}
    \theta_j = 0.17\left(\frac{1+z}{2}\right)^{-1/4}E_\mathrm{52}^{-1/4}A_\star^{1/4}t_j^{1/4},
\end{equation}
where $\theta_j$ is in radians, $E_\mathrm{52}$ is the isotropic-equivalent kinetic energy in units of $10^{52}$~erg, and $t_j$ is the jet break time in days. We calculate a lower limit of $\theta_j\gtrsim1.1$~deg, which is consistent with the sample range of $\theta_j=2.5\pm1.5$~degrees found by \citet{Wang2015}. The corresponding lower limit on the beaming correction is $f_b=(1-\cos\theta_j)\gtrsim1.8\times10^{-4}$, resulting in a beaming-corrected $\gamma$-ray energy of $E_\gamma\gtrsim1.7\times10^{50}$~erg and a kinetic energy of $E_\mathrm{K}\gtrsim5.2\times10^{50}$~erg. We calculate the radiative efficiency using $\eta_\gamma=E_{\gamma,\mathrm{iso}}/(E_{\gamma,\mathrm{iso}}+E_\mathrm{K,iso})$ and find $\eta_\gamma = 24.6$\%.

\begin{deluxetable}{lcc}
\tablecaption{Forward shock parameters\label{tab:MCMC}}
\tablehead{\colhead{Parameter} & \colhead{Highest-likelihood model} & \colhead{MCMC results}}
\startdata
$p$ & 2.27 & $2.28^{+0.01}_{-0.01}$\\
$E_\mathrm{K,iso}$ ($10^{53}$~erg) & 29.0 & $30.7^{+4.5}_{-3.1}$\\
$A_\star$ & $6.3\times10^{-3}$ & $4.7^{+1.6}_{-2.1}\times10^{-3}$\\
$\epsilon_e$ & $9.7\times10^{-3}$ & $8.7^{+1.3}_{-2.1}\times10^{-3}$\\
$\epsilon_B$ & $1.5\times10^{-1}$ & $2.4^{+3.7}_{-1.0}\times10^{-1}$ \\
$A_{V,\mathrm{host}}$ (mag) & 0.009 & $0.011^{+0.010}_{-0.007}$\\
$t_j$ (days) & $\gtrsim8.2$ & -\\
$\theta_j$ (deg) & $\gtrsim1.1$ & -\\
$E_\mathrm{K}$ ($10^{50}$~erg) & $\gtrsim5.2$ & - \\
$E_\gamma$ ($10^{50}$~erg) & $\gtrsim1.7$ & - \\
\enddata
\end{deluxetable}

\section{Explanations for the millimeter rebrightening}\label{sec:mm_rebrightening}
We consider a number of mechanisms that may give rise to a rebrightening at radio frequencies, including supernova emission, a counter jet, a density enhancement of the surrounding medium, interstellar scintillation, a two-component jet, energy injection, and reverse shock emission. 

\subsection{Supernova emission}
\citet{BarniolDuran2015} investigated the possibility of detecting radio emission from the highly-energetic Type Ic supernovae that are expected to accompany most long GRBs, and found that the emission from the shocked external medium swept up by the supernova blast wave should peak a few tens of years after the explosion, much later than our observed peak at 18.1 days. Furthermore, the luminosities of supernovae observed at millimeter wavelengths are generally 2--4 orders of magnitude fainter than those of GRB afterglows \citep{Eftekhari2022}. We therefore rule out supernova emission as a possible cause for the rebrightening in GRB 210702A. 

\subsection{Counter jet}
Once the GRB blast wave transitions into the non-relativistic regime, a bump in radio light curves may be expected from the receding counter jet, with simulations showing that the emission should peak more than $\sim1000$~days after the GRB \citep{Zhang2009}. The decelerating jet will experience a jet break prior to the non-relativistic transition. The lack of a jet break observed in our data therefore rules out a counter jet as a viable explanation. 

\subsection{Density enhancement}
Long GRBs offer a natural mechanism for producing a variable circumburst density since their progenitors are massive stars which produce stellar winds and undergo significant mass loss preceding their death. Several studies, however, have shown that even large density variations are unlikely to produce significant changes in the afterglow emission \citep{Nakar2007,Vaneerten2009,Gat2013,Geng2014}.

\subsection{Interstellar scintillation}\label{subsec:ISS}
Interstellar scintillation (ISS) due to inhomogeneities in the Galactic free electron density have been observed at radio wavelengths in a number of GRB afterglows \citep{Goodman1997, chandra2008,Vanderhorst2014,Greiner2018,Alexander2019}. \citet{Anderson2023} explored ISS as an explanation for the early, rapid variability in the radio light curves of GRB 210702A and found a transition frequency of $\nu_0=7.66$~GHz for the GRB line of sight. Our millimeter observations are more than an order of magnitude higher in frequency than this transition frequency and at a much later time so that the effects of ISS should be negligible. Our calculation of the modulation index of $m\lesssim2.5\%$ throughout our 97.5~GHz observations confirms that ISS is not relevant, though it may be relevant at lower frequencies (see Section \ref{subsec:discuss_FS_models}). We rule out ISS as a viable explanation for the millimeter rebrightening.

\subsection{Two-component jet}
A two-component jet has been invoked to explain the afterglows of several GRBs where a standard forward shock model does not suffice \citep{Berger2003,Racusin2008,Vanderhorst2014,Sato2023}. A faster, more narrowly-collimated inner jet gives rise to the early optical and X-ray emission while a slower, wider jet gives rise to later emission. If the rebrightening is due to forward shock emission from a wider jet, we would require a jet break from the inner jet near the start of the rebrightening around 8.1 days, otherwise a very steep rise of $\alpha\approx3$ (from our light curve fit in Section \ref{subsec:temporal}) would be required which is not possible from standard forward shock emission. No jet break is seen in our X-ray light curve, so this model appears to be disfavoured.

\subsection{Energy injection}
There are two physical forms of energy injection that may give rise to a rebrightening in afterglow light curves. The first involves a long-lasting central engine which injects a Poynting flux into the blast wave, such as a spinning-down millisecond magnetar \citep{Dai1998,Zhang2001a}. The luminosity of the central engine is usually given as a power-law in time beginning at a time $t_0$ and defined by the power-law index $q$: $L(t)=L_0(t/t_0)^{-q}$. The blast wave energy will only increase substantially when $q<1$, with the total energy increasing as $E_\mathrm{tot}\propto t^{1-q}$. The second form of energy injection invokes an impulsive central engine injection of a stratified ejecta with a power-law distribution in the bulk Lorentz factor: $M(>\gamma)\propto  \gamma^{-s}$ \citep{Rees1998,Sari2000}. It is impossible to physically distinguish the two scenarios from afterglow observations since both forms can be made equivalent through a relationship between the $q$ and $s$ parameters \citep{Zhang2006}. During the period of energy injection the light curves within each spectral regime will be altered according to the value of $q$ or $s$. Thereafter, the afterglow will evolve as a blast wave with the new, increased kinetic energy. 

We test the energy injection scenario by making use of the closure relations for energy injection within the $q$-formalism from \citet{Zhang2006}. At the start of the rebrightening the 97.5~GHz band is within the spectral regime with $\nu_m<\nu_\mathrm{97.5GHz}<\nu_c$. The appropriate closure relation in a wind medium is $\alpha=\frac{(2-2p)-(p+1)q}{4}$. Using our light curve rising index of $\alpha_\mathrm{97.5GHz,rise}=0.95$ and a value of $p=2.28\pm0.01$ derived from our theoretical modelling, we calculate a value of $q=-1.88$. The blast wave energy increases as $E_\mathrm{K,iso,1}=E_\mathrm{K,iso,0}\left(t_1/t_0\right)^{1-q}$ over the period of energy injection. We take $t_0$ and $t_1$ as the times corresponding to the start and peak of the rebrightening at 8.2 and 18.1 days, respectively. We find that the blast wave energy increases by a factor of $\approx10$ over the course of the rebrightening. 

\begin{figure*}
\centering
\includegraphics[width=\textwidth]{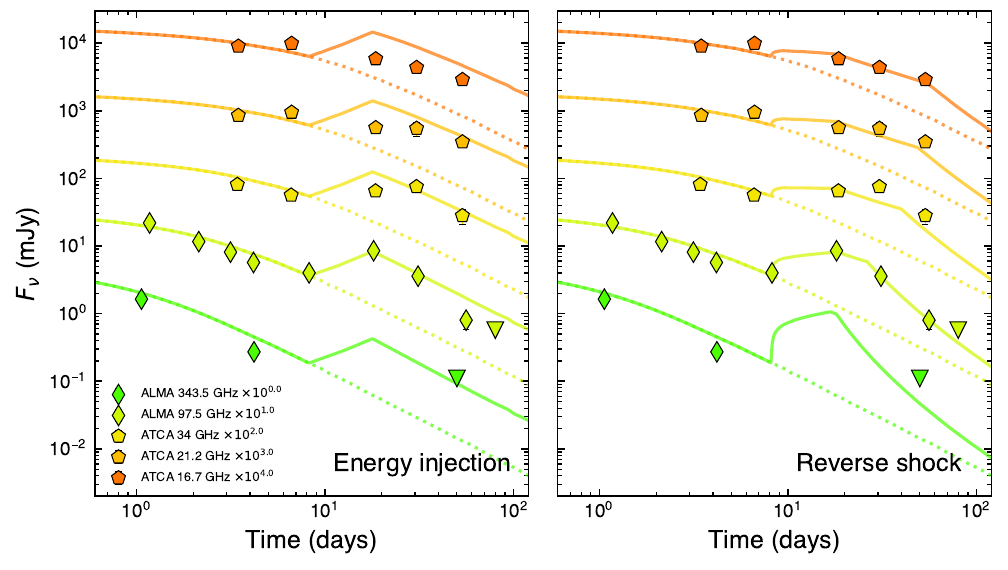}
\caption{Higher frequency ($\nu\geq16.7$~GHz) radio light curves with our energy injection (left) and reverse shock (right) models. The initial model (dotted line) is our highest-likelihood forward shock model from Section \ref{subsec:modelling}. We do not show the lower frequency ($\nu<16.7$~GHz) radio data since the initial forward shock model cannot explain these data (Figure \ref{fig:model_LCs}). We also do not show the optical and X-ray data as these data are unaffected by RS emission and energy injection at $t<10$~days.
} 
\label{fig:E_injection}
\end{figure*}

Taking our forward shock model from Section \ref{subsec:modelling} as the starting point, we introduce a period of energy injection in which the kinetic energy of the blast wave increases as $t^{1-q}$ between 8.2 and 18.1 days. We keep all other forward shock parameters the same. We exclude the lower frequency radio bands since there is no standard forward shock model prior to the rebrightening that can accommodate these data. The resulting model results in an achromatic rebrightening across all bands. The model can adequately capture the 97.5~GHz rebrightening, but does not fully capture the behaviour in the higher frequency bands (Figure \ref{fig:E_injection}). As discussed in Section \ref{subsec:radio}, however, possible unaccounted-for systematic uncertainties from the calibration of our ATCA data prevents us from completely excluding this scenario. 

\subsection{Reverse shock}
We now consider whether the 97.5~GHz rebrightening is caused by emission from a reverse shock (RS). There are two physical mechanisms that can produce RS emission: the first is a standard RS related to the forward shock (FS) that is produced by a shock wave moving back through the jet ejecta; the second is a RS produced by the late-time collision of a shell ejected by the GRB central engine. For both cases, the RS will emit synchrotron radiation characterised by its own set of spectral break
frequencies and peak flux: $\nu_{a,\mathrm{r}}$, $\nu_{m,\mathrm{r}}$, $\nu_{c,\mathrm{r}}$, and $F_{\nu_{m,\mathrm{r}}}$. For a standard RS, the RS and FS parameters are related at the deceleration (or shock crossing) time $t_\mathrm{dec}$, and can therefore be used to derive the ejecta Lorentz factor and magnetization. After the deceleration time, the flux above $\nu_{c,\mathrm{r}}$ is negligible since electrons are no longer accelerated within the ejecta. The evolution of the RS spectral breaks depends on whether the shock is relativistic or Newtonian in the frame of the unshocked ejecta. In the Newtonian case the spectral evolution depends on an additional parameter $g$ which dictates the profile of the shocked ejecta with $\gamma\propto r^{-g}$, and where we expect $1/2\lesssim  g\lesssim3/2$ from theoretical arguments \citep{Meszaros1999,Kobayashi2000}. 

If the 97.5~GHz rebrightening is caused by a standard RS related to the FS, our first optical detection at 0.004 days allows us set an upper limit on the deceleration time of $t_\mathrm{dec}<0.004$~days since we were able to model all of the optical emission with a FS (Section \ref{subsec:modelling}) which must have decelerated prior to this time. After shock crossing, it is only possible to obtain a rising and falling light curve through the passage of a spectral break, which must be $\nu_{a,\mathrm{r}}$ for a RS in a wind medium \citep{Zou2005}. Regardless of the ordering of the spectral breaks, we find that the RS emission in this case would outshine the detected optical emission at 0.004 days by several orders of magnitude, ruling out a standard RS scenario. 

Next we consider RS emission from the violent collision of two ejecta shells \citep{Zhang2002,Lyutikov2017,Lamberts2018}. Such a scenario has been used to model GRBs 140304A and 210672A \citep{Laskar2018a,Schroeder2023}. A requirement for a violent collision is that the second shell is moving much faster than the initial shell (which is responsible for the observed FS emission) at the collision time ($t_\mathrm{col}$), so that $\Gamma_2\gg\Gamma_1$. Furthermore, we rely on the assumptions that the energy of the initial shell does not increase enough to cause an increase in the FS emission and that FS emission from the second shell is negligible compared to the RS emission \citep{Zhang2002}. The RS emission from the second shell begins forming when the two shells collide at $t_\mathrm{col}$, and peaks once the RS has crossed the ejecta and fully decelerated at the deceleration time $t_\mathrm{dec}$. We choose to associate the collision time with the beginning of the rise in the 97.5~GHz light curve at 8.2 days, and the deceleration time with the peak in the light curve at 18.1 days. For a standard ordering of spectral breaks ($\nu_{a,\mathrm{r}}<\nu_{m,\mathrm{r}}<\nu_{c,\mathrm{r}}$), we find that a relativistic RS in which $\nu_{m,\mathrm{r}}$ crosses the 97.5~GHz band shortly after shock crossing can provide an adequate fit to our higher frequency radio data (Figure \ref{fig:E_injection}). At the shock crossing time, this model has $\nu_{m,\mathrm{r}}\approx 300$~GHz, $\nu_{a,\mathrm{r}}\lesssim16.7$~GHz, $\nu_{c,\mathrm{r}}\gtrsim10^{14}$~Hz, and $F_{\nu_{m,\mathrm{r}}}\approx1$~mJy. We note, however, that this is not a unique model since two of the break frequencies ($\nu_{a,\mathrm{r}}$, and $\nu_{c,\mathrm{r}}$) are not well-constrained. Futhermore, Newtonian RS models with both standard and non-standard orderings of spectral breaks can also accommodate the data. 

\section{Discussion}\label{sec:discussion}
\subsection{Radio rebrightenings}\label{subsec:mm_rebrightening}
Similar radio rebrightenings to the one observed in GRB 210702A have been observed in XRF 050416A and GRB 210726A \citep{Soderberg2007,Schroeder2023}. Radio observations of XRF 050416A at 1.43, 4.86 and 8.46~GHz showed an unusual flare peaking at $t\sim40$ days which was followed by rapid fading ($t^{-2}$), with the radio spectrum remaining optically thin throughout the observations. \citet{Soderberg2007} regarded a large circumburst density jump or late-time energy injection from a slow-moving shell as the most plausible explanation for the rebrightening. In the energy injection scenario, the decay following the peak should be the same as the asymptotic temporal decay prior to the injection. The fact that the post-peak decay rate of $\alpha\gtrsim-2$ was steeper than the predicted decay led \citet{Soderberg2007} to suggest that a jet break may have occurred on a similar timescale to the energy injection, a trend found by \citet{Laskar2015} in their study of energy injection in GRB afterglows. This scenario is in fact consistent with the post-peak decay rate in GRB 210702A of $\alpha_\mathrm{mm,decay}=-1.67\pm0.17$ being steeper than the predicted decay of $\alpha\approx-1.4$. 

More recently, \citet{Schroeder2023} studied the short GRB 210726A and found a radio flare at 6~GHz with a rapid rise between $\approx10-19$ days followed by a rapid decline at 3, 6 and 10~GHz. Energy injection and a reverse shock from a shell collision were regarded as plausible explanations to explain the late-time radio flare. Similar to the energy injection scenario for XRF 050416A, a jet break at the time of the peak of the flare was invoked to explain the steep post-peak temporal index of $\alpha=-2.1\pm0.4$. For the reverse shock shell collision scenario, an X-ray rebrightening at 4.6 days was attributed to the ejection of the shell while the peak of the radio flare was attributed to the deceleration time of the reverse shock. 

The most noteworthy difference between these previous studies and ours is that the observed rebrightening in these bursts occurred at centimeter wavelengths whereas the rebrightening in GRB 210702A occurred  in the millimeter regime, where scintillation is virtually guaranteed not to play a role (Section \ref{subsec:ISS}). Had millimeter data been obtained for these other bursts, it is possible that they may also have shown a rebrightening at millimeter wavelengths. Our observations demonstrate that millimeter light curves can exhibit some of the more complex features commonly seen at higher frequency optical and X-ray bands, and that these features may actually be more widespread than previous observations suggest, given that only a small number of bursts have millimeter follow-up light curves. 

\subsection{Failure of forward shock model}\label{subsec:discuss_FS_models}
Prior to the millimeter rebrightening, a standard forward shock model within a stellar wind medium was sufficient to account for the optical, early millimeter and late-time X-ray data in GRB 210702A. The early ($t<0.01$~day) X-ray data was not accounted for by our model, so we attributed the shallower early X-ray decline with $\alpha=-1.00\pm0.02$ to a period of energy injection. The greatest challenge to our model, however, was posed by the lower frequency ($\nu\leq21.2$~GHz) radio data. This includes multiple apparent spectral peaks in the SEDs and a flux suppression relative to the FS model that generally grows and persists for longer with decreasing observing frequency. For the former, one possible mechanism could be ISS, which was shown to be important in ATCA observations during the first day \citep{Anderson2023}. We now investigate this further. Using the ISS transition frequency ($\nu_0=7.66$~GHz) and scattering measure ($\mathrm{SM} = 1.06\times10^{-3}$~kpc\,m$^{-20/3}$) from \cite{Anderson2023} derived using RISS19 \citep{Goodman1997,Hancock2019}, we employ the prescription of \citet{Goodman2006} to calculate modulation indices (which represents fractional flux variability), and find that these range from $m\approx20\%$--$100\%$ at $\lesssim18$~days for frequencies between 1 and 22~GHz. While the predicted value of the modulation index decreases rapidly with both time and frequency, we nevertheless find $m\gtrsim50\%$ in the ATCA 5.5~GHz and 9~GHz bands for the full duration of the observations, indicating that at least some of the observed variations at these frequencies may be attributable to ISS. On the other hand, values of $m\lesssim10\%$ at 34~GHz and $m\lesssim2.5\%$ at 97.5~GHz confirm that the millimeter bands are unaffected by ISS and that the observed rebrightening is intrinsic to the source. 

For the latter (reduced observed flux relative to the model), we note that the flux suppression in the MeerKAT L band relative to the FS model spans a factor of 3--10 at 8.7--32~days, whereas we expect $m\lesssim40\%$ at 1.3~GHz during this period. As we discussed in Section \ref{subsec:radio}, this could be alleviated by requiring $1.28~{\rm GHz}\ll\nu_a\approx97.5~{\rm GHz}$ at 1.17~days. However, no standard model can accommodate such a high self-absorption frequency (Section \ref{subsec:modelling}). An additional means of suppressing the radio emission is to invoke a population of thermal, non-accelerated electrons within the shocked region which may lead to an increase in the self-absorption frequency by a factor of $\approx10-100$ \citep{Ressler2017,Warren2018}. Such an argument was made for GRB 210731A and GRB 210726A \citep{deWet2023a,Schroeder2023}. Nevertheless, the expected rapid evolution of $\nu_a$ in such a case would remain a challenge, and we defer further discussion of this model to future work.

Past sample studies of radio afterglows have shown that standard forward shock models are often incompatible with observations, which contrasts with observations at optical and X-ray afterglows that show general agreement with the forward shock models. \citet{Kangas2020} studied 21 GRBs with multi-wavelength data showing clear evidence for a jet break in their X-ray light curves and found that only half of the bursts had radio afterglows consistent with standard models. Most of the radio light curves were consistent with a single power-law, even after a jet break observed at higher frequencies. \citet{Levine2023} used standard afterglow closure relations to test whether radio observations of 84 GRBs were consistent with predicted temporal and spectral behaviour and, similar to \citet{Kangas2020}, found that only roughly half of the sample agreed with expectations. More recently, difficulties in modelling radio data have been encountered in the all time brightest GRB 221009A \citep{Bright2023,Laskar2023,Oconnor2023}. \citet{Laskar2023} found that the data could only partially be explained by a forward shock model, with additional emission components needed to explain the radio and millimeter emission, or nonstandard assumptions regarding the basic analytic models of relativistic synchrotron emission such as evolving microphysics parameters or an atypical electron energy distribution. \citet{Gill2023} instead modelled the broadband afterglow with a shallow angular structured jet in which all of the radio emission arises from a reverse shock rather than forward shock emission component. A two-component jet was proposed by other authors \citep{Sato2023,Zhang2023}. In summary, radio observations of GRB afterglows are highlighting the need for additional theoretical and modelling efforts in order to explain the data. We leave it to future work to test these alternative scenarios for GRB 210702A. 

\section{Conclusions}\label{sec:conclusion}
We have presented the results of our extensive multi-wavelength follow-up campaign of GRB 210702A, a bright long-duration GRB at $z=1.160$. Our 97.5~GHz ALMA light curve shows a rebrightening beginning at 8.2 days and peaking at 18.1 days, the first such rebrightening seen in a GRB millimeter light curve. Our X-ray, optical and millimeter data taken prior to the rebrightening can be explained by a standard forward shock model in a stellar wind medium with an electron energy spectral index of $p\approx2.27$. The shallow decay with $\alpha\approx-1$ seen in the optical light curve prior to 0.017 days and in the 97.5~GHz light curve prior to the rebrightening we attribute to the proximity of $\nu_m$ to the observing band. The shallow decay with $\alpha\approx-1$ in the X-rays, however, we attribute to a period of energy injection. Our radio light curves from 700~MHz to 34~GHz do not appear to evolve achromatically along with the millimeter rebrightening, though we cannot exclude this for the higher-frequency ATCA data due to inherent uncertainties arising from radio data calibration. Our radio SEDs show clear evidence for synchrotron self-absorbed emission, though we are unable to reconcile our constraints on the self-absorption break with any standard forward shock model. Our radio observations show similar incompatibilities found in modelling other GRBs, such as GRB 221009A \citep{Bright2023,Laskar2023,Oconnor2023}, and may require alternative scenarios such as a structured jet or thermal electron population. 

We considered a number of scenarios to explain the millimeter rebrightening in GRB 210702A and found that a period of energy injection into the forward shock or a reverse shock from a late-time shell collision are the most plausible mechanisms, yet both scenarios are unable to capture the behaviour seen in the other radio bands. Our observations demonstrate that millimeter light curves can display some of the more complex features seen in X-ray and optical light curves, such as rebrightenings or flares. Further millimeter observations will be required to determine the frequency of such features. 

\begin{acknowledgments}
The MeerLICHT consortium is a partnership between Radboud University, the University of Cape Town, the South African Astronomical Observatory (SAAO), the University of Oxford, the University of Manchester and the University of Amsterdam, in association with and, partly supported by, the South African Radio Astronomy Observatory (SARAO), the European Research Council and The Netherlands Research School for Astronomy (NOVA). We acknowledge the use of the Inter-University Institute for Data Intensive Astronomy (IDIA) data intensive research cloud for data processing. IDIA is a South African university partnership involving the University of Cape Town, the University of Pretoria and the University of the Western Cape. SdW and PJG are supported by NRF SARChI Grant 111692. RBD acknowledges support from the National Science Foundation under grant 2107932. TE is supported by NASA through the NASA Hubble Fellowship grant HST-HF2-51504.001-A awarded by the Space Telescope Science Institute, which is operated by the Association of Universities for Research in Astronomy, Inc., for NASA, under contract NAS5-26555. SB is supported by a Dutch Research Council (NWO) Veni Fellowship (VI.Veni.212.058). The research leading to these results has received funding from the European Union’s Horizon 2020 Programme under the AHEAD2020 project (grant agreement number 871158). This work made use of data supplied by the UK Swift Science Data Centre at the University of Leicester. The Australia Telescope Compact Array is part of the Australia Telescope National Facility\footnote{\href{https://ror.org/05qajvd42}{https://ror.org/05qajvd42}} which is funded by the Australian Government for operation as a National Facility managed by CSIRO. We acknowledge the Gomeroi people as the Traditional Owners of the Observatory site. We thank the staff of the GMRT that made these observations possible. The MeerKAT telescope is operated by the South African Radio Astronomy Observatory, which is a facility of the National Research Foundation, an agency of the Department of Science and Innovation. GMRT is run by the National Centre for Radio Astrophysics of the Tata Institute of Fundamental Research.  
\end{acknowledgments}

\bibliography{aas}{}
\bibliographystyle{aasjournal}

\appendix
\section{Table of flux measurements}

\startlongtable
\begin{deluxetable*}{cccccccc}
\tablewidth{0pt}
\tablecaption{GRB 210702A flux measurements \label{tab:flux}}
\tablehead{
\colhead{$\Delta t$ (days)} & \colhead{Telescope} & \colhead{Band/Filter} & \colhead{Frequency (Hz)} & \colhead{Flux (mJy)} & \colhead{Uncertainty (mJy)} & \colhead{Detection? ($1=$ yes)}
}

\startdata
0.00099 & \Swift{}/XRT & 1 keV & 2.42e+17 & 1.026390 & 0.129683 & 1 \\
0.00099 & \Swift{}/XRT & 1 keV & 2.42e+17 & 1.316908 & 0.163709 & 1 \\
0.00100 & \Swift{}/XRT & 1 keV & 2.42e+17 & 0.928455 & 0.114673 & 1 \\
0.00101 & \Swift{}/XRT & 1 keV & 2.42e+17 & 1.179080 & 0.160764 & 1 \\
0.00101 & \Swift{}/XRT & 1 keV & 2.42e+17 & 0.798587 & 0.209709 & 1 \\
\multicolumn{1}{c}{\ldots} & \multicolumn{2}{c}{\ldots} & \multicolumn{1}{c}{\ldots} & \multicolumn{2}{c}{\ldots} & \multicolumn{1}{c}{\ldots} \\
\tableline
0.00390 & \Swift{}/UVOT & $u$ & 8.652e+14 & 42.075 & 6.975 & 1 \\
0.01126 & MeerLICHT & $q$ & 5.169e+14 & 27.137 & 0.455 & 1 \\
0.01247 & MeerLICHT & $u$ & 7.889e+14 & 13.316 & 0.502 & 1 \\
0.01371 & MeerLICHT & $q$ & 5.169e+14 & 21.605 & 0.416 & 1 \\
\multicolumn{1}{c}{\ldots} & \multicolumn{2}{c}{\ldots} & \multicolumn{1}{c}{\ldots} & \multicolumn{2}{c}{\ldots} & \multicolumn{1}{c}{\ldots} \\
\tableline
0.496 & ATCA & 5.5 GHz & 5.50e+09 & 0.074 & 0.024 & 1 \\
0.530 & ATCA & 9 GHz & 9.00e+09 & 0.277 & 0.044 & 1 \\
0.550 & ATCA & 21.2 GHz & 2.12e+10 & 0.393 & 0.071 & 1 \\
0.550 & ATCA & 16.7 GHz & 1.67e+10 & 0.368 & 0.071 & 1 \\
1.061 & ALMA & 343.5 GHz & 3.44e+11 & 1.637 & 0.053 & 1 \\
\multicolumn{1}{c}{\ldots} & \multicolumn{2}{c}{\ldots} & \multicolumn{1}{c}{\ldots} & \multicolumn{2}{c}{\ldots} & \multicolumn{1}{c}{\ldots} \\
\enddata
\tablecomments{Table \ref{tab:flux} is published in its entirety in the machine-readable format. A portion is shown here for guidance regarding its form and content.}
\end{deluxetable*}
\end{document}